\documentstyle[prb,aps,epsfig]{revtex}

\begin{document}

\twocolumn

    \title{Conjugated donor-acceptor chromophores in solution:
	non-linearity at work}
    \author{Francesca Terenziani\thanks{Tel: +39 0521 905461;
	Fax: +39 0521 905556; E-mail: terenzia@nemo.unipr.it} 
	and Anna Painelli}
    \address{Dip. Chimica Generale ed Inorganica, Chimica Analitica 
	e Chimica Fisica, Universit\`a di Parma;
	INSTM--UdR Parma,
	Viale delle Scienze 17/A, I--43100 Parma, Italia}

\maketitle

\section{Introduction}

Conjugated donor-acceptor (DA) chromophores, also called push-pull 
chromophores, are molecules constituted by an electron-donor (D) and
an electron-acceptor (A) group connected by a $\pi$-conjugated backbone.
These molecules are widely investigated in several, apparently unrelated 
fields: they are common solvation probes~\cite{reichardt}, 
are molecules of choice for second order non-linear optics (NLO) 
applications~\cite{marder}, and are useful model systems for 
electron transfer~\cite{myers}.
All these applications exploit the presence of a low-lying excited state,
characterized by a different electronic distribution from the ground state.

Reactions involving charge migrations are 
ubiquitous in all fields of chemistry and
biochemistry, so that understanding their mechanism is
of special interest~\cite{zewail}.
These reactions often occur in solution, and are strongly influenced
by the medium itself, suggesting a strong coupling between solvent
and solute.
Nowadays, advances in laser technology allow to study the coupled reactant 
and solvent dynamics~\cite{zewail}.
In the typical experiment, a (series of) short pulse(s) of light
is used to perturb the electronic charge distribution of the solute;
the system is then interrogated by a retarded probe pulse.
Polar conjugated chromophores, due to their strongly 
solvent-dependent properties, are molecules of choice for this kind of 
experiments, offering as an additional bonus also the opportunity
to investigate the kinetics of electron transfer
in the absence of reactant diffusion~\cite{rossky}.
A proper interpretation of both steady-state and time-resolved experiments
is the key to gain important physical information.
However, standard interpretative schemes are based on linear perturbative
treatments of the solute-solvent interaction, so that they disregard one
of the most characteristic features of push-pull chromophores, i.e.
their large non-linearity.
We have already proposed a model that, accounting for the intrinsic
non-linearity of the electronic system, is able to rationalize steady-state
electronic and vibrational spectra of polar chromophores in solution.
In this paper the same model is extended to discuss time-resolved experiments.

\section{A model for non-linear interactions}

Since the properties of push-pull chromophores are dominated
by the lowest excited state, a two-state model describes
the essential physics governing the low-energy behavior of these molecules.
The electronic structure of push-pull chromophores can be described as
resonating between the fully neutral, $|DA\rangle$, and the zwitterionic,
$|D^+A^-\rangle$, states, separated by an energy $2z_0$ and mixed up
by the hopping integral, $-\sqrt{2}t$~\cite{oudar}. 
The ionicity operator, $\hat\rho$, measures the weight of $|D^+A^-\rangle$
in the ground- or excited-state. The expectation value of this operator,
$\rho =\langle\hat\rho\rangle$, corresponding to the polarity of the molecule
in the relevant state, only depends on the ratio $z_0/t$, so that the
electronic problem is fully described by a single parameter.
We allow for the coupling of the electronic system with a molecular
vibration, $Q$, and a solvation coordinate, $q$, by assigning the two
basis states two harmonic potential energy surfaces (PES), with equal
curvature but displaced minima along the two coordinates. 
The resulting Hamiltonian, describing these slow degrees 
of freedom and their coupling to electrons is~\cite{chemphys}:
\begin{equation}
{\mathcal {H}}' = \frac{1}{2} \Omega^2 Q^2 - 
	\sqrt{2\epsilon_{\rm{sp}}}\Omega Q\hat\rho +
	\frac{1}{2} \omega^2 q^2 - 
	\sqrt{2\epsilon_{\rm{or}}}\omega q\hat\rho
\end{equation}
where the first two terms account for the internal vibration, and 
the last two for the solvation coordinate.
$\Omega$ and $\omega$ are the frequencies corresponding to $Q$ and $q$,
respectively; $\epsilon_{\rm{sp}}$ and $\epsilon_{\rm{or}}$ are the energies
gained by $|D^+A^-\rangle$ due to the relaxation of $Q$ and $q$, respectively,
i.e. the small-polaron binding energy and the solvent reorganization
energy.

The linear dependence of the energy separation between the two basis states
on $Q$ and $q$ coordinates originates a non-linear dependence of 
$\rho$ on the two coordinates, making the exact potential energy, 
$V(Q,q)$, anharmonic. This anharmonicity,
originating from the coupling of electrons to slow degrees of freedom,
has many important consequences: in particular it is responsible for 
a large amplification of static hyperpolarizabilities~\cite{luca}.

\section{Spectroscopic properties}

Absorption and fluorescence spectra are vertical processes, occurring
at different $(Q,q)$. The dependence of $\rho$ on slow coordinates
(both $Q$ and $q$) then leads to non-trivial effects on absorption 
and fluorescence frequencies and band-shapes.
The vertical excited state reached upon photon absorption has
a different polarity with respect to the ground state.
Immediately after absorption, the slow degrees of freedom readjust 
in response to the new charge distribution of the solute.
But, as long as slow degrees of freedom relax, the solute molecule
itself feels a new surrounding and in turn readjusts its polarity.
The non-linear, self-consistent relaxation problem can be solved exactly
in our simple picture, to calculate the equilibrium polarity of the fully
relaxed excited state~\cite{cpl}.
Fluorescence is once more a vertical process that, starting from the
relaxed excited state, leads to the orthogonal ground state.
Our model then predicts different Huang-Rhys factors for absorption and 
fluorescence, and explains the observation of narrower 
fluorescence than absorption bands for
push-pull chromophores~\cite{cpl}.

The solvent orientational coordinate describes a very slow
(actually overdamped) motion and is responsible for inhomogeneous broadening 
effects in optical spectra.
This broadening can be modeled in terms of a statistical distribution 
of $q$ conformations, $w(q)$~\cite{jpca1}.
Since the chromophore readjusts its polarity in response to the solvent
configuration, one ends up with a distribution of chromophore polarities,
$w(\rho)$. 
Broadened spectra are then calculated as a sum of spectra corresponding
to different molecular polarities.

The self-consistent nature of the interactions implies that slow degrees
of freedom are in turn affected by the coupling to electrons.
The most apparent effects can be found in the dependence of vibrational
properties on the chromophore charge distribution and hence on the solvent
polarity. In agreement with experimental data, we predict 
solvent-dependent vibrational frequencies and inhomogeneous 
broadening of infrared and Raman bands in polar solvents~\cite{jpca1}.
Inhomogeneous broadening, affecting both electronic and vibrational states
in polar solvents, rationalizes an unusual phenomenon of 
dispersion of resonant Raman bands with the excitation line~\cite{jpca2},
observed for the dye phenol blue dissolved in polar solvents~\cite{yamaguchi}.

The dependence of molecular properties and hence of absorption and emission
band-shapes on the configuration of slow coordinates (in particular the 
solvation coordinate) is the key to understand some interesting phenomena
observed in time-resolved spectra.
Recent pump-probe and femtosecond hole-burning measurements on push-pull
chromophores dissolved in polar solvents~\cite{blanchard,ernsting}
have in fact revealed the appearance
of so-called temporary isosbestic points (TIPs) in transient 
gain and differential absorption spectra, respectively.
This means that spectra collected at different time delays seem
to cross in an isosbestic point, whose position actually depends on the 
chosen time window~\cite{ernsting}.
The physical origin of TIPs has not been understood so far. The only
temptative explanation~\cite{blanchard} invokes the active role of two 
(or more) excited states.
This explanation is not adequate, as proved by
the fact that TIPs do appear in transient
spectra collected for molecules dissolved in polar solvents only,
while this feature is not observed in non-dipolar solvents~\cite{blanchard}.
The appearance of TIPs in transient spectra is then related to
solvation dynamics.

In typical time-resolved experiments, the relaxation of internal molecular 
vibrations is in general completed after the first few hundreds of 
femtosecond following the excitation~\cite{ernsting}, so that subsequent 
dynamics can be ascribed to the relaxation of the solvation coordinate 
toward the new equilibrium configuration.
Then, in the framework of our model, we can calculate the 
time evolution of transient spectra based on the temporal evolution of the 
probability distribution $w(\rho)$, which, in turn, can be obtained 
from the evolution of $w(q)$ on the relevant PES. 
In the phase space of coordinate and momentum $(q,p)$ the evolution of 
$w(q,p)$ can be described by the Fokker-Planck equation. 
In the overdamped regime
(friction coefficient, $\gamma$, greater than $\omega$), relevant to
the solvation coordinate, the Fokker-Planck equation reduces to
the simpler Smoluchowski equation, in the $q$-space only~\cite{mukamel}:
\begin{equation}
	\frac{\partial w(q)}{\partial t} = \frac{1}{\gamma} 
	\left[ w(q)\frac{\partial^2 V}{\partial q^2} + 
	\frac{\partial V}{\partial q} \frac{\partial w(q)}{\partial q} + 
	kT\frac{\partial^2 w(q)}{\partial q^2} \right]
\end{equation}
where $V$ is the anharmonic potential energy for the relevant state, 
$k$ the Boltzmann constant, $T$ the temperature.
An example of the temporal evolution of $w(q)$ on one of the PES is given 
in Fig.~1(a) from $t=0$, when an 
out-of-equilibrium population is created, to $t \rightarrow \infty$, 
when equilibrium is reached.
In Fig.~1(b) the corresponding evolution of $w(\rho)$ is reported:
due to anharmonicity and to the non-linear relation between $\rho$ and $q$,
the distribution is asymmetric and evolves in width and shape.

Transient properties only depend on $\gamma$ and $\omega$ via
the ratio $\tau=\gamma / \omega^2$.
In order not to add any adjustable parameter to the original model,
we fix $\tau$ to the longitudinal relaxation time of the pure solvent,
as obtained from literature data.
In Fig.~2(a), we show transient emission spectra, calculated
for the parameters reported in the caption. The evolution
of the band-shape, due to the dependence of electronic properties on the 
configuration of slow variables, together with the red-shift,
is responsible for the appearance of a TIP. 
These data reproduce the gain spectra measured for the polyene PA2 
dissolved in dioxane~\cite{blanchard}. 
The adopted $\tau$ value coincides with the accepted value of the
longitudinal relaxation time of dioxane~\cite{blanchard}.
Fig.~2(b) refers to a different experiment. In particular, to simulate
femtosecond hole-burning spectra of coumarin 102 (C102) dissolved in 
CH$_3$CN~\cite{ernsting}, we have calculated
transient differential absorption spectra
from an out-of-equilibrium $q$-distribution in the ground state, 
as obtained by a pump-dump sequence. The spectra, calculated for the
model parameters that fit steady-state spectra of C102~\cite{prepa},
and for $\tau=0.3$~ps (the accepted value for the longitudinal relaxation
time of CH$_3$CN~\cite{blanchard}) are in good agreement with 
experimental data.
Specifically, the blue-shift of the transient absorption is accompanied
by an evolution of the absorption band-shape, so that spectra collected
in a narrow temporal window all cross in a unique point.
This point is not truly isosbestic and, in agreement with experimental 
data, we calculate a slow blue-shift of its position at
increasing time.

\section{Conclusions}

The proposed model describes the spectral properties of conjugated 
donor-acceptor chromophores in solution in terms of two electronic states
coupled to internal vibrations and to solvent degrees of freedom.
The model is simple enough to allow for a full exploitation of the
non-linearity of the electronic system, that manifests itself in a
self-consistent interplay between electronic, vibrational and solvation
degrees of freedom. Signatures of this non-linearity
are recognized in several typical spectral features that have not explanation
in standard approaches. The model is semiempirical
in nature: the few model parameters are fixed against experimental data.
Then the applicability of the model itself and the reliability of the
model parameters for a specific molecule must be confirmed via
the comparison with a large body of experimental data, much wider than
the pool of data used to fix the parameters.
In this respect we underline that the scope of our model is extremely wide,
ranging from steady-state to time-resolved spectra, from electronic to
vibrational data. Based on available experimental data, the model has been
validated for the phenol blue dye, whose steady-state electronic
and vibrational spectra are extensively discussed in Ref.~\cite{jpca2}.
A second interesting case study is offered by C102, whose
time-resolved electronic spectra have been discussed in the previous
Section. For this molecule, solvation effects have also been observed
in steady-state and time-resolved vibrational spectra~\cite{elsaesser}: 
also these effects can be rationalized within our model~\cite{prepa}.
Linear electron-phonon and
solute-solvent interactions originate, in materials with non-linear 
responses, large and non-trivial effects that cannot be understood based
on perturbative treatments.

\acknowledgments

Work supported by the Italian National Research Council (CNR)
within its ``Progetto Finalizzato Materiali Speciali per Tecnologie
Avanzate II'', and by the Ministry of University and of Technological 
and Scientific Research (MURST).

\vfill
\eject

\onecolumn

\begin{figure}
\vspace{5cm}
\begin{center}
\mbox{\epsfig{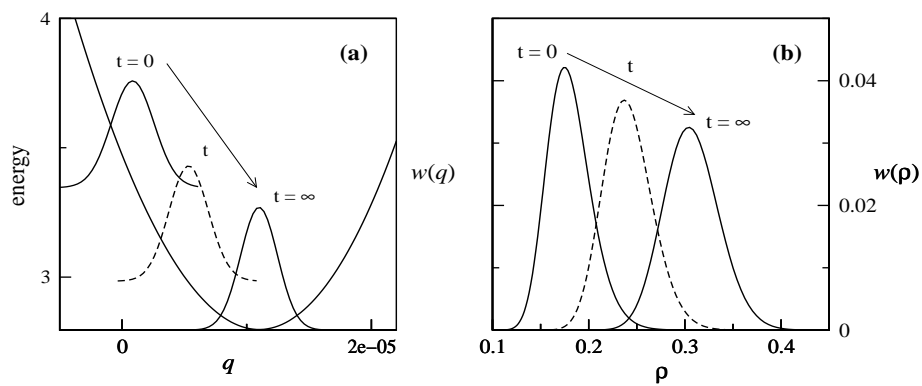}}
\end{center}
\caption{
(a) Evolution of the probability distribution $w(q)$ 
over the represented PES.
(b) Corresponding evolution of the probability distribution $w(\rho)$.}
\end{figure}

\vfill
\eject

\begin{figure}
\begin{center}
\mbox{\epsfig{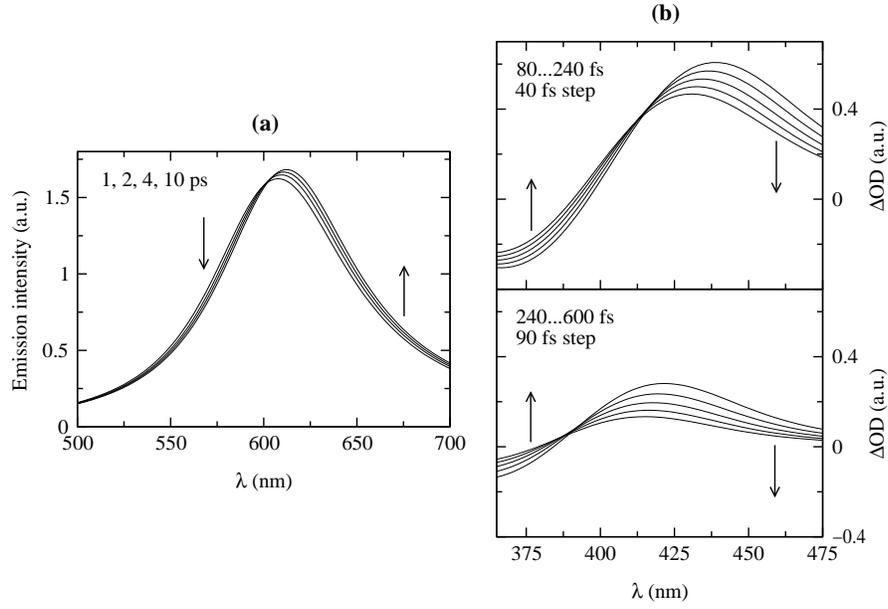}}
\end{center}
\caption{
(a) Transient emission spectra calculated for $z_0 = 0.5$~eV, 
$\sqrt{2}t=1$~eV, $\Omega=0.2$~eV, $\epsilon\rm{_{sp}}=0.22$~eV, 
$\epsilon\rm{_{or}}=0.4$~eV, $\tau=1.7$~ps, $T=300$~K. 
Time delays are displayed on the figure, time increasing
in the direction of the arrows.
(b) Transient differential absorption spectra calculated for 
$z_0 = 1$~eV, $\sqrt{2}t=1.2$~eV, $\Omega=0.2$~eV, 
$\epsilon\rm{_{sp}}=0.3$~eV, $\epsilon\rm{_{or}}=0.35$~eV,
$\tau=0.3$~ps, $T=300$~K. 
Upper panel and lower panel refer to different time windows,
as labeled on the figure. Time increases in the direction of the arrows.}
\end{figure}

\end{document}